\documentstyle[prl,aps,preprint]{revtex}

\tighten

\begin{document}
\newcommand{\dy}{$^{152}$Dy}
\newcommand{\J}{{\cal J}^{(2)}}
\newcommand{\ep}{\epsilon}
\newcommand{\bra}[1]{\langle #1 |}
\newcommand{\ket}[1]{| #1 \rangle}
\renewcommand{\labelenumi}{{(\roman{enumi})}}

\draft

\title{Octupole Correlations in Excited Bands of Superdeformed \dy}
\author{Takashi~Nakatsukasa\thanks{Present address:
AECL Research, Chalk River Laboratories, Chalk River,
Ontario K0J 1J0 Canada}}
\address{Research Center for Nuclear Physics, Osaka University,
Ibaraki 567, Osaka, Japan}
\author{Kenichi~Matsuyanagi}
\address{Department of Physics, Kyoto University, Kyoto 606--01, Japan}
\author{Shoujirou~Mizutori}
\address{
Department of Mathematical Physics, Lund Institute of Technology,
Box 118, S-22100\\ Lund, Sweden}
\author{Witold~Nazarewicz\thanks{On
leave of absence from
Institute of Theoretical Physics,
Warsaw University, Warsaw, Poland;
Institute of Physics, Warsaw University of Technology,
Warsaw, Poland.}
}
\address {Joint Institute for Heavy-Ion Research,
Physics Division\\
Oak Ridge National Laboratory,
P.O. Box 2008, Oak Ridge, TN 37831 U.S.A.\\
and\\
Department of Physics, University of Tennessee,
Knoxville, TN 37996 U.S.A.
}

\maketitle

\bigskip
\begin{abstract}
RPA calculations, based on the cranked shell model, are performed
for superdeformed \dy\
in which five excited bands have been found recently.
We show that characteristic features of the observed
dynamical moments of inertia are well accounted for by explicitly
taking the octupole correlations
into account.
Importance of the interplay between rotation and octupole
vibrations is stressed, and it is suggested that one of the
observed  excited
bands might be a collective octupole vibration built on
the superdeformed yrast band.
\end{abstract}

\newpage

The discovery of superdeformed (SD) rotational bands has opened
many new avenues in studies of nuclear structure at the extremes
of rapid rotation and large deformation.
Recent  experimental developments, especially large  $\gamma$-ray
detector  arrays  (Eurogam,  Gammasphere, Ga.Sp,  etc.),
have  offered
better observational limits which help in  clarifying many
aspects of high-spin nuclear structure.

Recently, five excited SD bands (Bands $2 - 6$) have been
observed in \dy\ in a Eurogam experiment\cite{Dag94}.
According to various theoretical calculations\cite{BAR88,NWJ89,SVB90},
\dy\  has
a SD doubly-closed-shell configuration corresponding
to the large single-particle gaps  at  Z=66 and N=86.
Since  the pairing correlations in SD bands in the A=150  region  are
expected to be seriously quenched
due to the low level-density of single-particle states
and rapid rotation,  the angular momentum variations in the  dynamical
moments  of inertia, ${\J}=dI/d\omega$, are mainly due to
the intrinsic angular momentum alignment
of single-particle  orbitals,
especially high-$N$ intruder orbitals\cite{BAR88,NWJ89,SVB90}.
Consequently, the $\J$ moments of inertia carry
important  experimental information about
single-particle configurations in SD bands.

The excited SD bands in \dy,
observed by Dagnall et al.\cite{Dag94},
have a very low intensity relative to
the yrast SD band.
This might be related to the predicted  SD magic structure in {\dy}.
Due to its magic structure,
collective excitation modes
are expected to influence the properties of
near-yrast  SD bands in {\dy}. In this context, octupole vibrations play
a very special role.
According to the RPA calculations based on
the cranked shell model\cite{MSM90,NMM92},
low-lying octupole vibrations are more important than
low-lying quadrupole vibrations built on the SD
shape.
Strong octupole correlations in SD states
have also been suggested theoretically in Refs.
\cite{DWS90,Abe90,HA90,Cul90,Bon91,LDR91,ND92,Ska92,Ska93}.
The calculations of Ref.\cite{NMM93}
demonstrate that the inclusion of the
coupling between
quasiparticle and octupole vibrational modes is important
for understanding the experimental data for SD $^{193}$Hg\cite{Cul90}.

In this Letter, we discuss octupole correlations in excited SD bands
of \dy.
Comparing our results with the experimental data, we propose a
plausible scenario
for the microscopic structure of  excited SD bands in {\dy}.
This scenario is compatible with the discussions by Dagnall et
al.\cite{Dag94} but the influence of octupole correlations is
explicitly considered.
Indeed, one of the excited SD bands is suggested to
have a collective octupole vibrational character.
If this interpretation is correct, this is the first case
in which the collective vibrational mode
at SD high-spin states has been observed
experimentally.\footnote{Recently,
an excited band in  SD $^{190}$Hg
has been interpreted in terms of
octupole vibrations\cite{Cro94}.}

In order to investigate the influence of octupole vibrations
 on the
excitation spectrum of SD \dy, the RPA treatment has
been carried out. The model
Hamiltonian has been assumed to be of the form:
\begin{equation}
\label{hamiltonian}
H= h'_{\rm s.p.}
-\frac{1}{2} \sum_K \chi_{3K} Q_{3K}^{''\dagger} Q_{3K}^{''}
-\frac{1}{2} \sum_K \chi_{1K} (\tau_3 D_{1K})^{''\dagger}
 (\tau_3 D_{1K})^{''}\ ,
\end{equation}
where $h'_{\rm s.p.}$ is a cranked single-particle Nilsson Hamiltonian,
$h'_{\rm s.p.}=h_{\rm Nilsson} -\omega_{\rm rot}\hat{J_x}$,
and $Q_{3K}^{''}=(r^3Y_{3K})^{''}$ and
$D_{1K}^{''}=(rY_{1K})^{''}$ are, respectively,  the
doubly stretched octupole and dipole operators
defined by coordinates
$x_i^{''}=\frac{\omega_i}{\omega_0}x_i$\cite{SK89}.
The equilibrium quadrupole deformations have been determined by means of the
shell correction method. A large configuration space composed of
nine major shells for both protons and neutrons has been used for solving the
coupled RPA dispersion equations.
The spurious velocity dependence associated with the $l^2$ and
$\vec l\cdot \vec s$ terms in the Nilsson potential
are removed by means of the method
proposed in  Ref. \cite{Kin88}.
We note that the obtained single-particle routhians are similar
to those for the Woods-Saxon potential\cite{NWJ89}.
The pairing gaps $\Delta_p$ and $\Delta_n$ are assumed to be zero:
Although dynamical pairing fluctuations never vanish, {\it relative}
energy spectra and {\it relative} alignments are known to be well
described by the simple cranked shell-model routhians without pairing
at $\omega_{\rm rot}\geq 0.3$\,MeV/$\hbar$, i.e., in the region
where the experimental data are available\cite{Shi90,Rag93}.
In order to determine the isoscalar coupling strengths, $\chi_{3K}$,
we have carried out the systematic RPA calculations
 for the low-frequency $I^\pi=3^-$
states in medium-heavy nuclei. Guided by these calculations,
we use
$\chi_{3K}=1.05\chi_{3K}^{\rm HO}$
where $\chi_{3K}^{\rm HO}$ are the selfconsistent values for the harmonic
oscillator potential\cite{SK89}.
For the isovector dipole coupling strengths we use
$ \chi_{1K}=-{\pi V_1}/{\langle (r^2)^{''}\rangle } $
with $V_1=140$\,MeV\cite{BM75}.

Figure~\ref{omega}(a) shows the RPA eigenvalues
as functions of rotational frequency
$\omega_{\rm rot}$.
The lowest excitation mode with signature $\alpha=1$ (dotted line) can
be associated with the collective octupole vibrational band.
The band has $K=0$ at $\omega_{\rm rot}=0$,
but the $K$-mixing due to the
Coriolis force is significant at high
rotational frequencies.
The B(E3)-values calculated at $\omega_{\rm rot}=0$
in the strong coupling scheme are around
$\mbox{B(E3; } 3^- \rightarrow 0^+ )\approx  35\mbox{ W.u.}$
By comparing Fig.~\ref{omega}(b) and (c), we see that the
octupole collectivity carried by the lowest $\alpha=1$ band decreases gradually
with $\omega_{\rm rot}$.
On the other hand, collectivity of the lowest excitation mode with
$\alpha=0$ (solid line) is weak and this mode has a dominant 1p-1h
configuration at high frequency.
Excitation energy of this band drastically decreases
in the high-frequency region and its
alignment, $i=-\frac{dE_x}{d\omega}$, is evaluated to be about $5\hbar$.
Since this band has much lower excitation energy at high frequency
than the octupole vibrational $\alpha=1$ band,
it may be populated with higher intensity.

Calculations show that
the neutron N=86 single-particle shell gap persists at high
frequencies,
while the proton Z=66 shell gap vanishes at high angular momenta
where the proton $N=7$ ($\alpha=-1/2$)  orbital
crosses the fourth $N=6$ ($\alpha=-1/2$) orbital
(see Fig.~\ref{routhian} and discussion
in Ref.\cite{Dag94}).
The 1p-1h excitation associated with these two orbitals gives rise to
the lowest excited state with signature $\alpha=0$.
The alignment of this 1p-1h excitation is equal to
$i_{\rm p}-i_{\rm h}\approx 4.5\hbar$; i.e.,
the large alignment of the band comes from the intrinsic angular momentum
of the proton intruder $N=7$ orbital.
In contrast, the lowest 1p-1h excitation with $\alpha=1$ is
associated with the proton $N=7$ ($\alpha=-1/2$)
and the third $N=6$
($\alpha=1/2$) orbital. Its
excitation energy is about 1~MeV higher than that of the
$\alpha=0$ band in the highest frequency region.
Because of this effective energy gap, the collective
mode with $\alpha=1$ survives up to rather high frequencies.
Since the alignment of the collective octupole phonon is
less than $3\hbar$, the lowest $\alpha=0$ band  carries a larger alignment
and becomes lower at high frequency.

In the following, we discuss the dynamical moments of inertia of
Bands 2, 3, and 6 for which octupole correlations are calculated to
be important.
Characteristic features of
Bands 2, 3, and 6, determined in Ref.\cite{Dag94},
can be summarized as follows:
(i) $\J$ of Band 2 (Band 3) has a bump (dip) at $\omega_{\rm rot}
\approx 0.5$MeV/$\hbar$;
(ii) Bands 2 and 3 are populated with higher intensity compared to
other excited bands (Bands $4 - 6$);
(iii) $\J$ of Band 6 is larger than that of the SD yrast band
and is almost constant as a function of rotational frequency;
(iv) At low values of
$\omega_{\rm rot}$ Band 6 shows a decay branch into the yrast SD band.

On the basis of the above observations,
we propose a scenario
in which the lowest and
the second lowest excited $\alpha=0$ states (solid lines
in Fig.~\ref{omega}), and the lowest $\alpha=1$ state (dotted line)
correspond to Bands 2, 3, and 6, respectively.
Firstly, the $\J$ bump in Band 2 and the dip of Band 3
occurring at the same
frequency can be associated with crossing between the two lowest
$\alpha=0$ states, see Fig.~\ref{omega}.
Secondly, the high intensity of Bands 2 and 3 indicates
that at high frequency these bands have lower excitation energy than
the other bands.
Our conjecture is consistent with the intensity data for Band 2.\footnote{
On the other hand, calculations suggest that intensity of Band 3 should
be weaker than that of
Band 2 whereas experimentally it is similar; this weakens
our interpretation of Band 3.}
Thirdly, weak $\omega_{\rm rot}$-dependence of $\J$ in Band 6
suggests an almost constant curvature
$\frac{d^2E_x}{d\omega^2}$ of the routhian (see eq.~(\ref{j2})).
Finally, the partial decay of Band 6 into the yrast SD band
indicates that Band 6 may be a collective band possessing significant
(E1) transition matrix elements into the yrast SD band.

In order to make the comparison
with experimental data quantitative,
we calculate the dynamical moments of inertia $\J$.
They can be decomposed as
\begin{equation}
\label{j2}
\J=\J_0 + \frac{di}{d\omega}
= \J_0 - \frac{d^2E_ x}{d\omega^2}\ ,
\end{equation}
where $\J_0$ denotes the dynamical moment of inertia of the yrast
SD band of \dy\ (RPA vacuum).
We approximate the experimental $\J_0$ by the
Harris expansion,
$\J_0=\alpha + \beta\omega^2$, with
$\alpha=88.5\hbar^2\mbox{MeV}^{-1}$ and
$\beta=-11.9\hbar^4\mbox{MeV}^{-3}$.
Calculated and experimental values of $\J$ are compared in Fig.~\ref{moi};
it is seen that the characteristic features of the experimental data
are well reproduced.
It is worth noting that the octupole correlations are also important for
reproducing experimental $\J$ values for Bands 2 and 3.

In order to discuss the collectivity of octupole correlations,
we show in Fig.~\ref{wf}
the forward RPA amplitudes\footnote{
Sums of the squared backward RPA amplitudes,
$\sum_{\alpha\beta} |\varphi_n(\alpha\beta)|^2$,
at $\omega_{\rm rot}=0.3$MeV/$\hbar$ are 0.13, 0.11 and 0.48 for Bands
2, 3 and 6, respectively.}
$\psi_n(\alpha\beta)$
for Bands 2, 3, and 6.
We see that Bands 2 and 3 correspond to simple 1p-1h
excitations at the highest frequency region; {i.e.},
proton $N=6 \rightarrow N=7$
and proton $N=6 \rightarrow N=5$ excitations,
respectively.
Bands 2 and 3 cross
at $\omega_{\rm rot}^{(c)}\approx 0.5$\ MeV/$\hbar$.
For $\omega_{\rm rot} < \omega_{\rm rot}^{(c)}$,
collective components in both bands are significant.
In fact, the interaction matrix element between Bands 2 and 3
would be too small to reproduce the observed bumps and dips of $\J$
if octupole correlations were turned off.
On the other hand, Band 6 has vibrational character
in the whole range of rotational frequency.
The octupole collectivity of this band decreases with rotational frequency.

In summary, we have investigated the
effects of octupole correlations
in excited SD bands of \dy\
by means of the RPA based on the cranked shell model.
We found that a low-lying octupole vibrational band ($\alpha=1$)
appears near the yrast band ($E_x\approx 1\ $MeV).
According to our
scenario,
Bands 2, 3, and 6 have negative parity.
Band 2 (3) is the
lowest (second lowest) $\alpha=0$ band.
Band 6 is
the octupole vibrational $\alpha=1$ band.
The collectivity of Band 6 is expected to
gradually decrease with
$\omega_{\rm rot}$, while Bands 2 and 3 cross each other
at $\omega_{\rm rot}\approx 0.5\ $MeV/$\hbar$.
The calculated $\J$ values reflect the $\omega_{\rm rot}$-dependence
of the internal structures of these bands, and seem to agree
well with major characteristics found experimentally.

We would like to thank C.W.~Beausang and P.J.~Twin
for valuable discussions.
We are indebted to the ECT*
for financial support which
made possible
our stay at the ECT* during the International Workshop on
High Spins and Novel Deformations.
Computational calculations were supported in part by the RCNP,
Osaka University, as a RCNP Computational Nuclear Physics Project
(Project No. 93--B--02).
This work was also supported in part by the U.S. Department of Energy through
Contract Numbers DE-AC05-840R21400, DE-FG07-87ER40361, and DE-FG05-93ER40770.

\begin{figure}[pht]
\caption{\protect\small
Results of RPA calculations at
quadrupole deformation $\delta_{\rm osc}=0.59$.
(a) Calculated RPA eigenvalues (in MeV) for SD \dy, plotted as
functions of rotational frequency $\omega_{\rm rot}$ (in MeV/$\hbar$).
Solid (dotted) lines indicate negative-parity states with signature
$\alpha=0$ ($\alpha=1$).
The lowest $\alpha=1$ state
has $K=0$ in the limit $\omega_{\rm rot}=0$.
(b) Electric octupole strength
$\sum_K |\bra{n} \frac{1}{2}(1+\tau_3)Q_{3K}\ket{0}|^2$
at $\omega_{\rm rot}=0.3$MeV/$\hbar$ \ in Weisskopf units
($\ket{0}$ and $\ket{n}$ denote the RPA ground state and excited
states, respectively).
Solid and dotted lines indicate the $\alpha=0$ and $\alpha=1$ states,
respectively.
The vertical axis represents the excitation energy as in (a).
(c) The same as (b), except for $\omega_{\rm rot}=0.6\,$MeV/$\hbar$.
}
\label{omega}
\end{figure}

\begin{figure}[pht]

\caption{\protect\small
Neutron and proton single-particle routhians
as functions of
rotational frequency $\omega_{\rm rot}$.
The Nilsson parameters ($\kappa, \mu$) are adopted
from ref. \protect\cite{Nil69},
and the spurious velocity dependence
associated with $l^2$ and $\vec l\cdot
\vec s$ terms are removed according to a prescription developed by
Kinouchi and Kishimoto\protect\cite{Kin88}.
Orbitals having
parity and signature, $(\pi,\alpha)=(+,1/2)$, $(+,-1/2)$, $(-,1/2)$,
and $(-,-1/2)$ are shown by solid, dashed, dotted, and dash-dotted
lines, respectively.
The oscillator quantum number, $N_{\rm osc}$, is
indicated for ``high-$N$'' orbitals.
}
\label{routhian}
\end{figure}

\begin{figure}[pht]
\caption{\protect\small
Calculated (solid lines) and experimental (symbols)
dynamical moments of
inertia for excited SD bands (Bands 2, 3, and 6) in $^{152}$Dy.
Dotted lines indicate $\J$ for the yrast SD band,
which is approximated by the Harris formula
$\J_0=\alpha + \beta\omega^2$ with
$\alpha=88.5\hbar^2\mbox{MeV}^{-1}$ and
$\beta=-11.9\hbar^4\mbox{MeV}^{-3}$. See text for details.
}
\label{moi}
\end{figure}
\begin{figure}[pht]
\caption{\protect\small
Absolute values of the forward amplitudes, $|\psi_n(\alpha\beta)|$,
of the lowest
and the second lowest RPA solutions in the $\alpha=0$ sector
(portions a and b),
and the lowest RPA solution in the $\alpha=1$ sector (portion c),
corresponding to Bands 2, 3, and 6, respectively.
Solid (Dashed) lines indicate neutron (proton) amplitudes.
All amplitudes whose absolute values are greater than $1.5\times 10^{-1}$
are displayed. The characteristic p-h excitations  are indicated.
}
\label{wf}
\end{figure}

\end{document}